\documentclass{article}
\usepackage{cite}
\usepackage{booktabs}
\usepackage{algorithmic}
\usepackage{graphicx}
\usepackage{textcomp}
\usepackage{xcolor}
\usepackage{gensymb}
\usepackage{pgfplots}
\usepackage{tikz}
\usetikzlibrary{plotmarks}
\usepackage{soul}  % to strike through for comments
\usepackage[normalem]{ulem}
\pgfplotsset{compat=1.11}
\newlength\fwidth

\usepackage{calligra}
\usepackage{mathtools,xparse}
\usepackage{appendix}
\usepackage{eucal}
\usepackage{amsmath,amsfonts,amssymb,amsthm,amssymb}
\usepackage{bbm}
\usepackage{commath}
\usepackage[mode=buildnew]{standalone}% requires -shell-escape

\usepackage[preprint]{spconf}
\ninept

\title{Misalignment Recognition in Acoustic Sensor Networks Using a Semi-Supervised Source Estimation Method and Markov Random Fields}

\twoauthors
 {Gabriel F Miller, Andreas Brendel, Walter Kellermann}
	{Friedrich-Alexander-Universität, Erlangen-Nürnberg\\
	Multimedia Communications and Signal Processing Lab\\
	Cauerstr. 7, D-91058 Erlangen, Germany\thanks{This work was partially funded by the Deutsche Forschungsgemeinschaft (DFG, German Research Foundation) -- 282835863 -- within the Research Unit FOR2457 ``Acoustic Sensor Networks'' and from the European Union's Horizon 2020 Research and Innovation Programme, Grant Agreement No.~-871245.
    }}
 {Sharon Gannot}
	{Bar-Ilan University\\
	Faculty of Engineering\\
	Ramat-Gan, Israel}

\begin{document}

\maketitle

\begin{abstract}
In this paper, we consider the problem of acoustic source localization by acoustic sensor networks (ASNs) using a promising, learning-based technique that adapts to the acoustic environment. In particular, we look at the scenario when a node in the ASN is displaced from its position during training. As the mismatch between the ASN used for learning the localization model and the one after a node displacement leads to erroneous position estimates, a displacement has to be detected and the displaced nodes need to be identified. We propose a method that considers the disparity in position estimates made by leave-one-node-out (LONO) sub-networks and uses a Markov random field (MRF) framework to infer the probability of each LONO position estimate being aligned, misaligned or unreliable while accounting for the noise inherent to the estimator. This probabilistic approach is advantageous over na\"{i}ve detection methods, as it outputs a normalized value that encapsulates conditional information provided by each LONO sub-network on whether the reading is in misalignment with the overall network. Experimental results confirm that the performance of the proposed method is consistent in identifying compromised nodes in various acoustic conditions.  
\end{abstract}


\section{Introduction}

Sound source localization is a topic that has been covered in great detail and remains a burgeoning field of study \cite{BLG_speaekerTrackingMultipleMani, RS_music, RR_esprit, CK_generalizedCorrelationMethodforTimeDelay, AD_2dSoundsourceLocalBinMani, AD_VariEMforBinSoundSepAndLocal, AD_acousticSpaceLearningforSoundSep,TM_probModelRobustLocalizationBinauralAudioFront,BLG_rtfModelingSupervisedSourceLocalization, LG_semisupervisedSourceLocalization}, see \cite{9079214} for an overview of the state of the art. Especially, smart-home technology drove the need for robust and efficient localization methods in acoustic sensor networks (ASNs) \cite{cobos_survey_2017,griffin_localizing_2015}. While in the past, traditional localization methods typically relied on physics-based models \cite{RS_music,RR_esprit,CK_generalizedCorrelationMethodforTimeDelay}, there has been a growing interest in localizing acoustic sources using learning-based methods whereby position estimates are obtained directly from previously learned knowledge about a given acoustic environment. These methods have been shown to be effective, particularly in adverse acoustic conditions \cite{BLG_speaekerTrackingMultipleMani, LG_semisupervisedSourceLocalization, AD_2dSoundsourceLocalBinMani, AD_VariEMforBinSoundSepAndLocal, AD_acousticSpaceLearningforSoundSep,brendel_distributed_2019} as long as the parameters used for training remain static. For example, when localizing sources in a smart-home environment, many of the underlying characteristics of the room remain essentially unchanged (e.g., the room dimensions and reverberation time). This means the variability regarding the acoustic transfer functions, which are typically represented in a high-dimensional feature space, can be mostly attributed to the source position. This lends credibility to the use of learning-based methods where these static qualities can be captured during a training phase.

Due to the difficulty in acquiring labelled data, a semi-supervised method based on a small labelled and a large unlabelled data set is generally employed. Unlabelled data, which are easy to obtain, are used together with a few labelled `anchors' to train models for acoustic source localization \cite{BLG_rtfModelingSupervisedSourceLocalization}.
In \cite{BLG_RTFModSuperSource}, a semi-supervised approach was employed for source localization using a relative transfer function (RTF)-based feature vector, which measures the relation between the acoustic paths from a sound source to two different microphones. 
% The authors assert that the semi-supervised method is beneficial for source localization as the amount of labelled data available is limited to the acoustic environment being considered. 
Thus, by leveraging unlabelled data, a more robust localizer is achieved. In this study the scenario considered was limited to a single microphone system in a static environment with white Gaussian noise input.
Subsequently, in \cite{LG_semisupervisedSourceLocalization}, the semi-supervised inference approach, based on Gaussian process (SSGP) on multiple manifolds, was further developed and adapted to localize a speech source using a multi-microphone system, again based on a dense grid of RTFs \cite{RC_maniGPReg,LG_semisupervisedSourceLocalization,MB_semiSupervisedLocalizationDeepGenMod}. 

\begin{figure}
    \setlength\fwidth{0.42\textwidth}
    \centering
\begin{tikzpicture}

\begin{axis}[%
width=0.951\fwidth,
height=0.5\fwidth,
at={(0\fwidth,0\fwidth)},
scale only axis,
xmin=0,
xmax=3,
xlabel style={font=\color{white!15!black}},
xlabel={Shift Size (m)},
ymin=0.1,
ymax=0.5,
ylabel style={font=\color{white!15!black}},
ylabel={Estimation Error ($\mathrm{m}$) After Shift},
axis background/.style={fill=white},
xmajorgrids,
ymajorgrids,
legend style={at={(0,0.85)}, anchor=south west, legend cell align=left, align=left, draw=white!15!black, font=\footnotesize}, legend columns = 2
]
\addplot [color=red, dashed, line width = 1pt]
  table[row sep=crcr]{%
0	0.1352\\
0.05	0.1352\\
0.25	0.1352\\
0.45	0.1352\\
0.65	0.1352\\
0.85	0.1352\\
1.05	0.1352\\
1.25	0.1352\\
1.45	0.1352\\
1.65	0.1352\\
1.85	0.1352\\
2.05	0.1352\\
2.25	0.1352\\
2.45	0.1352\\
2.65	0.1352\\
2.85	0.1352\\
3.05	0.1352\\
};
\addlegendentry{Static ($T_{60} = 0.2\,\mathrm{s}$)}

\addplot [color=red, line width = 1pt]
  table[row sep=crcr]{%
0	0.1352\\
0.05	0.2103\\
0.25	0.2454\\
0.45	0.2137\\
0.65	0.226\\
0.85	0.1687\\
1.05	0.2667\\
1.25	0.2624\\
1.45	0.2204\\
1.65	0.2284\\
1.85	0.2427\\
2.05	0.247\\
2.25	0.2526\\
2.45	0.2253\\
2.65	0.2024\\
2.85	0.2573\\
3.05	0.2927\\
};
\addlegendentry{Dynamic ($T_{60} = 0.2\,\mathrm{s}$)}

\addplot [color=blue, dashed, line width = 1pt]
  table[row sep=crcr]{%
0	0.241\\
0.05	0.241\\
0.25	0.241\\
0.45	0.241\\
0.65	0.241\\
0.85	0.241\\
1.05	0.241\\
1.25	0.241\\
1.45	0.241\\
1.65	0.241\\
1.85	0.241\\
2.05	0.241\\
2.25	0.241\\
2.45	0.241\\
2.65	0.241\\
2.85	0.241\\
3.05	0.241\\
};
\addlegendentry{Static ($T_{60} = 0.4\,\mathrm{s}$)}

\addplot [color=blue, line width = 1pt]
  table[row sep=crcr]{%
0	0.241\\
0.05	0.3265\\
0.25	0.3728\\
0.45	0.3059\\
0.65	0.3161\\
0.85	0.3351\\
1.05	0.3801\\
1.25	0.3373\\
1.45	0.3403\\
1.65	0.3341\\
1.85	0.3432\\
2.05	0.3588\\
2.25	0.3525\\
2.45	0.3493\\
2.65	0.3569\\
2.85	0.3715\\
3.05	0.3162\\
};
\addlegendentry{Dynamic ($T_{60} = 0.4\,\mathrm{s}$)}

\addplot [color=green, dashed, line width = 1pt]
  table[row sep=crcr]{%
0	0.2728\\
0.05	0.2728\\
0.25	0.2728\\
0.45	0.2728\\
0.65	0.2728\\
0.85	0.2728\\
1.05	0.2728\\
1.25	0.2728\\
1.45	0.2728\\
1.65	0.2728\\
1.85	0.2728\\
2.05	0.2728\\
2.25	0.2728\\
2.45	0.2728\\
2.65	0.2728\\
2.85	0.2728\\
3.05	0.2728\\
};
\addlegendentry{Static ($T_{60} = 0.6\,\mathrm{s}$)}

\addplot [color=green, line width = 1pt]
  table[row sep=crcr]{%
0	0.2728\\
0.05	0.3631\\
0.25	0.3842\\
0.45	0.4216\\
0.65	0.3092\\
0.85	0.3596\\
1.05	0.3878\\
1.25	0.3718\\
1.45	0.3627\\
1.65	0.369\\
1.85	0.4043\\
2.05	0.3263\\
2.25	0.3902\\
2.45	0.3591\\
2.65	0.2855\\
2.85	0.3616\\
3.05	0.3822\\
};
\addlegendentry{Dynamic ($T_{60} = 0.6\,\mathrm{s}$)}

\end{axis}
\end{tikzpicture}%
    \caption{Mean error of the SSGP localizer for different reverberation times $T_{60}$, comparing scenarios without node movement  (dotted lines) to scenarios when one node in the network moved (random node shifted from its learned position) as can be seen in solid lines. }
    \label{moveError}
\end{figure}
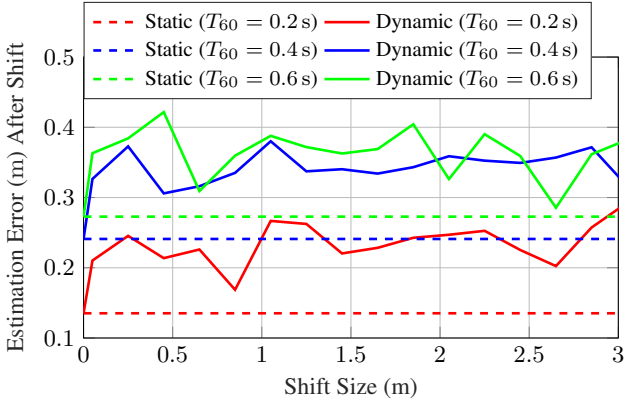

However, if the array constellation, e.g., the position of one or more nodes, changes relative to the training stage, the usefulness of the learned model becomes uncertain.
In our work, we adopt the SSGP method and consider the scenario where any given microphone node can be moved. The detrimental effect of an array movement on the localization error can be observed in Fig.~\ref{moveError}, where the error almost doubles with only a small shift of a random node in the network. We are thus posed with the problem of determining if a node is moving, and specifically determining which of the nodes is moving. In order to address both issues, we consider a technique recently introduced in the field of robotics for recognizing sensor misalignment \cite{NA_misalignmentRecMRFsFcn}. The authors in \cite{NA_misalignmentRecMRFsFcn} utilize Markov random fields (MRFs) with fully connected latent variables (FCLVs) to measure the probability of misalignment of a sensor network based on individual sensor readings and a ground truth mapping of a given room \cite{CB_patternRec,PK_efficientInferenceFCCRFs}. Recognition of misalignment is needed in \cite{NA_misalignmentRecMRFsFcn} to determine whether differences in measurements sampled over time should be attributed to actual changes or due to inherent noise. 

Rather than taking each sensor signal independently, for our scenario we look at the so-called leave-one-node-out (LONO) sub-network position estimates (with each sub-network containing all but one node) obtained via the SSGP method. We then use the differences between the position estimates before and after movement of a single node as input to the MRF model (note, for our considerations in this paper the sound source is static). Eventually our model outputs posterior probabilities per LONO sub-network for belonging to one of the following latent states: aligned, misaligned or unreliable. These posteriors are used to indicate both the probability of movement in the network, and also allows for inference of which node moved. 

%The remainder of this paper proceeds as follows. In Sec.~\ref{ssgpTech} we briefly review the SSGP approach to localization and in Sec.~\ref{misDetect} we present the MRF detection model. We then discuss the parameters of our simulation as well as the results of the detector compared to a na\"{i}ve estimate in Sec.~\ref{procedural_overview} , and in Sec.~\ref{conclusion} we conclude with a summary of our findings, as well as an outlook on potential future work. 

\section{Review of the SSGP source localization technique} \label{ssgpTech}

We now briefly review the SSGP source localization method (see \cite{LG_semisupervisedSourceLocalization} for details) and consider a speech signal in the Short-Time Fourier Transform (STFT) domain, $S\left(\tau,k\right)$, at frame index $\tau$, frequency index $k$, received at a given node $m$, and emitted from position $\mathbf{q}$. We then model the signal received at node $m$ as follows:
\begin{equation} \label{micSig_eqn}
        Y_i^m\left(\tau,k\right) = A_i^m\left(\tau,k,\mathbf{q}\right)S\left(\tau,k\right)+U_i^m\left(\tau,k\right)
\end{equation} 
with $i$ specifying the $i$th microphone in the $m$th node. 
%Additionally, $S(\tau,k)$ is the STFT representation of a sound originating at $\mathbf{q}$, \textit{Y$_i^m$}$\left(\tau,k\right)$ is the STFT representation of the signal received at, microphone \textit{i}, node \textit{m}, 
Additionally, $A_i^m\left(\tau,k,\mathbf{q}\right)$ is the acoustic transfer function (ATF) relating the sound source originating at position $\mathbf{q}$ to the $i$th microphone, and $U_i^m\left(\tau,k\right)$ is the STFT-domain representation of an additive noise signal which corrupts the measurement. Obviously, the spatial information required for localizing a source at position $\mathbf{q}$ is embedded in the ATF, and is independent of the source signal. Rather than extracting the ATF we use the aforementioned RTF feature vector, $\mathbf{h}^m$ (defined as the ratio of two ATFs \cite{SG_signalEnhancementUsingBeamformingAndNonstationarity}) as it is easier to acquire in practice and is equally informative for the proposed localization method.

In order to determine the position of an unknown source, we first define $\mathbf{q}_t = \left[q_{t,x}, q_{t,y}, q_{t,z}\right]^\top$ as the unknown `test' position to be inferred given some unknown RTF sample, $\mathbf{h}^m_t$, which relates the unknown source position $\mathbf{q}_t$ to node $m$, assuming that each of the $M$ network nodes has only two microphones. For training the SSGP estimator, a set of $n_D$ sound sources is used where $n_D$ is the number of training points, from which $n_U$ are measured RTFs and $n_L$ are measured RTFs with associated source positions serving as labels ($n_L+n_U=n_D$). Each Cartesian coordinate $p_{d,a}$, $a\in\{x,y,z\}$ of a training position, $\mathbf{p}_d\in\mathbb{R}^{3\times 1}$, is said to be the output of some target function, $f_a^m\left(\mathbf{h}^m_d\right)$ which relates the training position, $\mathbf{p}_d$, to node $m$ via an RTF sample $\mathbf{h}^m_d$. Moreover, we assume that the coordinates of all $n_D$ labelled and unlabelled training positions captured by vectors $\mathbf{p}_{D,a} = \left[p_{1,a}\dots,p_{n_D,a}\right]^\top$, are each jointly Gaussian, and the target functions, $f_a^m$, can be alternatively defined as the posterior mean function of corresponding Gaussian distributions. We will now discuss how we utilize the RTF training samples $\{\mathbf{h}^m_d\}_{d=1}^{n_D} \left(\forall m \in \{1,\dots,M\}\right)$ in order to identify the position of an unknown source from its corresponding RTF $\mathbf{h}^m_t$. We will omit the dependency on the coordinate $a\in\{x,y,z\}$ in the following for conciseness.

\begin{figure}
    \setlength\fwidth{0.55\textwidth}
    \centering
    \input{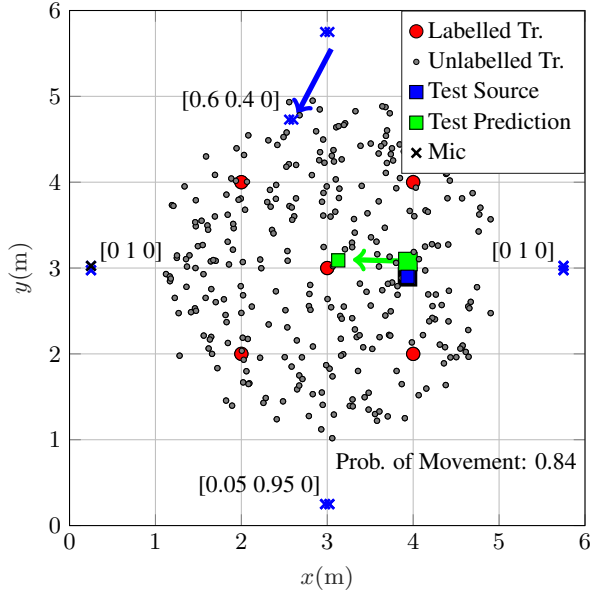}
    \caption{Misaligned scenario with example of proposed detection method. Green arrow indicates how the prediction of an acoustic source changes based on the movement and rotation of a random node (blue arrow). Values in brackets next to nodes indicate probability of a LONO sub-network being aligned, misaligned, or unreliable where the referenced node is the one left out. Higher probabilities of alignment indicate the node left out is likely compromised.}
    \label{misalignedEx}
\end{figure}

In order to localize a sound source, RTFs obtained at each node are compared to those obtained at every other node in the ASN. 
The relation between all RTFs are summarized via the kernel-based covariance matrix, $\pmb{\Sigma}_L$, with each element representing a pairwise affinity between two RTF samples. In particular, we express a given element in the covariance matrix which relates two labelled source positions, $l_i$ and $l_j$ as follows:
\begin{align}
        \left(\pmb{\Sigma}_L\right)_{l_i,l_j} 
        %&= \mathrm{cov}(\mathbf{p}_{l_i}, \mathbf{p}_{l_j}) \equiv \tilde{k}\left(\mathbf{h}_{l_i}, \mathbf{h}_{l_j}\right) \label{cov_element}\\
        &= \frac{1}{M^2} \sum_{d=1}^{n_D}\sum_{q=1}^{M}\sum_{w=1}^{M} k_q\left(\mathbf{h}_{{l}_i}^q, \mathbf{h}_d^q\right)k_w\left(\mathbf{h}_{l_j}^w, \mathbf{h}_d^w\right).
\end{align}
Here, $\mathbf{h}_{l_{i}}$, $\mathbf{h}_{l_{j}}$ are RTF samples from the set of labelled RTFs $\mathcal{H}_L = \{\mathbf{h}_{l_{i}}\}^{n_L}_{i=1}$, and $k_m(\mathbf{h}_i^{m}$, $\mathbf{h}_j^{m})$ is a conventional pairwise Gaussian kernel function, $k_m: \mathcal{M}_m \times \mathcal{M}_m \rightarrow \mathbb{R}_+$ with:
\begin{equation} \label{gauss_kern}
        k_m\left(\mathbf{h}_i^{m}, \mathbf{h}_j^{m}\right) = \exp\left\{-\frac{\norm{\mathbf{h}_i^{m} - \mathbf{h}_j^{m}}_2^2}{\varepsilon_m}\right\}
\end{equation} 
where $\mathcal{M}_m$ denotes a manifold corresponding to node $m$, and $\varepsilon_m$ is a parameter defining the width of the kernel \cite{MG_classesOfKernels}.
Similarly, we can define an element in the test covariance vector, $\pmb{\Sigma}_{Lt} \in \mathbb{R}^{n_{L}\times 1}$, used for inferring the position of an unknown source element: 
\begin{equation} \label{cov_element}
        {\left({\pmb{\Sigma}}_{Lt}\right)}_{l_i} = \frac{1}{M^2} \sum_{d=1}^{n_D}\sum_{q=1}^{M}\sum_{w=1}^{M} k_q\left(\mathbf{h}_{l_i}^q, \mathbf{h}_d^q\right)k_w\left(\mathbf{h}^w_t, \mathbf{h}_d^w\right).
\end{equation} 
The unknown position, $\mathbf{q}_t$, can thus be estimated coordinate-wise via the conditional mean with respect to the corresponding multivariate Gaussian distribution, $\mathbbm{P}\left(q_{t}\mid\mathbf{p}_{L}, \mathcal{H}_L\right)$, where $\mathbf{p}_{L} \in \mathbb{R}^{n_L\times 1}$ is the vector containing only a coordinate of the labelled training positions. The distribution of all source positions (known and unknown) is defined over the concatenation of all coordinates of the labelled training positions $\mathbf{p}_{L}$ and the coordinate to be estimated $q_{t}$:
\begin{equation} \label{concat_pos_est}
    \begin{bmatrix}
        \mathbf{p}_{L}\\
        q_{t}
    \end{bmatrix}\bigg\vert \mathcal{H}_L \backsim \mathcal{N}\left(\mathbf{0}_{n_{L+1}}, 
    \begin{bmatrix}
        \pmb{\Sigma}_L + \sigma^2\mathbf{I}_{n_L} & \pmb{\Sigma}_{Lt}\\
        \pmb{\Sigma}_{Lt}^\top & \Sigma_t
    \end{bmatrix}
    \right)
\end{equation} 
where $\Sigma_t$ is the variance of $q_{t}$, $\sigma^2$ is the variance associated with the accuracy of the labels, $\mathbf{I}_{n_L}$ is the $n_L \times n_L$ identity matrix and $\mathbf{0}_{n_{L+1}}$ is an all-zero vector of length $n_{L+1}$. 
Thus, we estimate the position of an unknown source by the conditional mean associated with \eqref{concat_pos_est}:
\begin{equation}\label{test_est}
        q_{t} = \mu_\mathrm{cond} = %\pmb{\Sigma}_{Lt}^\top\hspace{0.1cm}\tilde{\pmb{p}}_L
        \pmb{\Sigma}_{Lt}^\top\hspace{0.1cm}\left(\pmb{\Sigma}_L + \sigma^2\mathbf{I}_{n_L}\right)^{-1}\mathbf{p}_L.
\end{equation} 
%with \hspace{.03cm} $\tilde{\pmb{p}}_L = \left(\pmb{\Sigma}_L + \sigma^2\mathbf{I}_{n_L}\right)^{-1}\mathbf{p}_L$ being a set of weights independent of any test source. 
An example of the localization scenario is shown in Fig.~\ref{misalignedEx} (detailed room specifications can be found in Sec.~\ref{procedural_overview}.).

%\begin{figure}
    %\setlength\fwidth{0.42\textwidth}
    %\centering
    %%\includegraphics[width=5.75cm,height=6cm]{figures/alignedEx.JPG}
    %\input{figures/alignedEx.tex}
    %\caption{Room setup with application example of the SSGP source localization technique (distance is in meters).}
    %\label{alignedEx}
%\end{figure}
\section{Misalignment detection} \label{misDetect}
%MRFs provide a convenient and consistent way of modeling context dependent entities and can be implemented in a local and massively parallel manner \cite{CB_patternRec,LI_mrfModelingCompVis}.

In our scenario, MRFs are used to determine if a node in an ASN is displaced, and also determine which node moved. MRFs provide a convenient and consistent way of modeling context dependent entities and can be implemented in a local and massively parallel manner \cite{CB_patternRec,LI_mrfModelingCompVis}. MRFs are especially useful for inference if a priori probability functions are given for the latent variables governing the observations. The observed quantities we use as input to the MRF model are the SSGP localization estimates of $M$ LONO sub-networks. These estimates are compared to the localization estimates recorded before movement for the $m$th LONO sub-network via the Euclidean distance 
\begin{equation}\label{error_def}
        %e_m=\sqrt{\sum_{i\in\{x,y,z\}}\left((\mathbf{q}_{m})_i-(\hat{\mathbf{q}}_{m})_i\right)^2}.
        e_m = \Vert \mathbf{q}_{m} - \hat{\mathbf{q}}_{m}\Vert_2.
\end{equation}
Note that, here, $\mathbf{q}_{m}$ refers to position estimates recorded by a given LONO sub-network before movement occurs, and $\hat{\mathbf{q}}_{m}$ refers to the estimate after movement. While in dynamic scenarios with moving sources these distinctions will not be so clear, in the static scenario assumed here, they are useful for the desired analysis.
The latent variables are given by the errors made by a LONO in a given acoustic environment, which is dependent on the room itself and, consequently, on the variability of the SSGP localizer. Therefore, in practice, the considered MRF model (as detailed in Sections 3.1, 3.2) compares the difference obtained from each LONO sub-network using a message passing scheme and incorporates prior information regarding the general localization error distribution. This distribution is acquired by simulating localization estimates of LONO sub-networks after a random array in the ASN is shifted in a random direction with random rotation and comparing it with the ground truth position of the source. The output of the model is a probability indicating if the network is in alignment. We assume the latent variables to be FCLVs to ensure the difference in estimation measured by each LONO sub-network is compared with the difference measured by every other LONO sub-network. Additionally, we assume that only one node in the network is moving at a time, therefore, the sub-network with the smallest probability of movement as determined by the MRF model would probably be the one that did not contain the moved node, thus allowing us to infer which particular node was moving. For this inference, we consider the posterior probability output by the MRF that a given LONO sub-network $m$, is of one of the following latent classes: aligned, misaligned or unreliable.  
    
\subsection{Likelihood distributions of estimation errors}
For approximating the latent posterior probabilities, $\mathbbm{P}\left(\mathbf{z}_m\mid\mathbf{e}\right)$, where $\mathbf{z}_m=\left[z_{m,1}, z_{m,2}, z_{m,3}\right]^\top$ is an indicator vector of binary variables with each variable indicating whether a given LONO sub-network belongs to a given latent class, and $\mathbf{e}$ is the vector containing the difference in estimates from all LONO sub-networks.

We first define the prior (error) distributions of each latent class, which were found empirically from observed errors. For the aligned case, $\mathbf{z}_m=\left[1,0,0\right]$, we choose a half-normal distribution with variance $\sigma_\mathrm{align}^2$ \cite{NA_misalignmentRecMRFsFcn,RHB_halfNormalDistr}. 
\begin{equation}
    \mathbbm{P}\left(e_m\mid \mathbf{z}_m=\left[1,0,0\right], \sigma^2_\mathrm{align}\right) = 2\mathcal{N}\left(e_m; 0,\sigma^2_\mathrm{align} \right),\label{prior distributions} e_m \geq 0,
\end{equation}
an exponential distribution with parameter $\lambda$ for the misaligned case, $\mathbf{z}_m=\left[0,1,0\right]$
\begin{equation}
    \mathbbm{P}\left(e_m\mid \mathbf{z}_m=\left[0,1,0\right], \lambda\right) = \frac{\lambda\exp\{-\lambda\,e_m\}}{1 - \exp\{-\lambda\,e_\textrm{max}\}},
\end{equation}
and a uniform distribution for unreliable observations, $\mathbf{z}_m=\left[0,0,1\right]$
\begin{equation}
    \mathbbm{P}\left(e_m\mid \mathbf{z}_m=\left[0,0,1\right]\right) = \mathrm{unif}\left(0,e_\mathrm{max}\right),
\end{equation}
where, $e_\mathrm{max}$ references the maximum localization error. A uniform distribution is assigned to the unreliable class (analogous to the assumption made in \cite{NA_misalignmentRecMRFsFcn}) to reflect the uninformative character of this class, as we assume that the movement in the network cannot be predicted.

\subsection{Latent class estimation and failure detection}
As noted, we make the FCLV assumption which allows us to consider the viewpoint of every LONO sub-network in calculating the set of latent posterior probabilities for a specific LONO sub-network. In particular, every set of latent variables associated with a LONO sub-network receives messages from all other nodes and their corresponding set of variables to initialize the marginal posterior probabilities, which is calculated as
\begin{equation} \label{fclv_initPosterior}
    \begin{aligned}
        \mathbbm{P}\left(\mathbf{z}_m\mid\mathbf{e}\right)=\frac{1}{Z}\mathbf{l}_m\odot\prod_{\substack{m'=1 \\ m'\neq m}}^{M} \pmb{\mu}_{m' \rightarrow{m}}\left(\mathbf{z}_m\right).
    \end{aligned}
\end{equation} 
Here, $Z$ is a normalizing factor, $\odot$ is the Hadamard product and $\mathbf{l}_m$ is a likelihood vector
\begin{equation} \label{likelihood_vector}
        \mathbf{l}_m = \left[\mathbbm{P}\left(e_m\mid z_{m,1}\right), \mathbbm{P}\left(e_m\mid z_{m,2}\right), \mathbbm{P}\left(e_m\mid z_{m,3}\right)\right]^\top.
\end{equation} 
The message from $m'$th to the $m$th LONO sub-network is denoted as
\begin{equation} \label{fclv_initMsg}
        \pmb{\mu}_{m' \rightarrow m }\left(\mathbf{z}_m \right) = \psi _{m',m}\left(\mathbf{z}_{m'}, \mathbf{z}_m\right)\mathbf{l}_{m'}.
\end{equation} 
In this case, $\psi _{m',m}\left(\mathbf{z}_{m'}, \mathbf{z}_m\right)$ is the transition probability from state $m'$ to $m$ and is an element of the transition matrix $\pmb{\psi} \in \mathbb{R}_+^{3\times3}$. The matrix is optimized using an iterative proportional fitting procedure  based on empirical localization errors \cite{CB_patternRec,SF_IPFP}. 

Finally, after each node receives initial messages from all other nodes, messages are continually passed around until convergence to the maximum likelihood posterior.

%\subsection{Localization failure detection}
With the posteriors for each LONO sub-network, we obtain the probability of misalignment in the overall network based on the average posterior probabilities of misalignment for all sub-networks:

\begin{equation} \label{p_failure}
        p_\mathrm{failure} = \frac{1}{M}\sum^M_{m=1}\mathbbm{P}\left(z_{m,2}\mid\mathbf{e}\right).
\end{equation} 
Then, the criterion $p_\mathrm{failure} \geq p_\mathrm{thresh}$ with the user-defined threshold $p_\mathrm{thresh}$ is used for detecting node movement. A misaligned scenario is illustrated in Fig.~\ref{misalignedEx} where a node is displaced by one meter. Values in brackets next to each node indicate the probability of a LONO sub-network being aligned, misaligned, or unreliable where the referenced node is the one left out. Higher probabilities of alignment indicate that the node left out is more probable to have moved.

\section{Evaluation} \label{procedural_overview}
We present a simulation study showing the efficacy of the proposed method. After describing the experimental setup we discuss the results obtained from Monte-Carlo simulations.

We consider a room of size $6\,\mathrm{m} \times 6\,\mathrm{m} \times 3\,\mathrm{m}$ with four nodes uniformly spaced in a square (see Fig.~\ref{misalignedEx}), each comprising two microphones spaced $5\,\mathrm{cm}$ apart. The Region of Interest (RoI) was chosen to be in the center of the node network and within a $2\,\mathrm{m}$ radius from the center of the room. In total, we simulated five labelled sources and 300 unlabelled sources to generate RTFs, whereby each unlabelled point was randomly chosen from a uniform 2D distribution within the RoI. White noise convolved with simulated room impulse responses (RIRs) \cite{EH_rirGen} has been used for training.

The SSGP parameters were optimized via an ML estimation (see \cite{LG_semisupervisedSourceLocalization} for details) for varying noise levels and $T_{60}$. This was done by drawing at random speech signals from a database of English speakers \cite{speakerData}, again convolving them with simulated RIRs, randomizing the position of the source and comparing the positional estimates to the ground truth position. The parameters of the MRF model, $\sigma_\mathrm{align}^2$, $\lambda$, and $e_\mathrm{max}$ were chosen via a random grid search whereby a room environment was simulated and arrays were randomly shifted \cite{KE_stochasticOptViaGridSearch}. The optimal parameters were then chosen based on the detector's ability to recognize movement for a range of probability thresholds. Care was taken in choosing these thresholds, as extremely small thresholds result in a high number of false positives as even a movement occurring with only a small probability will be declared movement by the MRF model, and without loss of generality, large thresholds result in a large number of false negatives. Thus the threshold was incremented (from 0 to 1 by increments of 0.05 m) to balance the range of possible outcomes. 
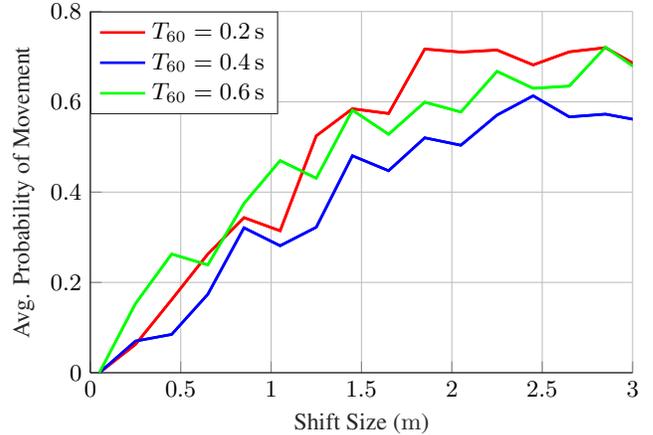
\begin{figure}
    \setlength\fwidth{0.45\textwidth}
    \centering
\begin{tikzpicture}

\begin{axis}[%
width=0.9\fwidth,
height=0.6\fwidth,
at={(0\fwidth,0\fwidth)},
scale only axis,
xmin=0,
xmax=3,
xlabel style={font=\color{white!15!black}},
xlabel={Shift Size ($\mathrm{m}$)},
ymin=0,
ymax=0.8,
ylabel style={font=\color{white!15!black}, align=center},
ylabel={Avg. Probability of Movement},
axis background/.style={fill=white},
axis x line*=bottom,
axis y line*=left,
xmajorgrids,
ymajorgrids,
legend style={at={(0,1)}, anchor=north west,legend cell align=left, align=left, draw=white!15!black}
]
\addplot [color=red, forget plot, line width = 1pt]
  table[row sep=crcr]{%
0.05	0\\
0.25	0.06268\\
0.45	0.1618\\
0.65	0.2628\\
0.85	0.3435\\
1.05	0.3143\\
1.25	0.5246\\
1.45	0.5846\\
1.65	0.5741\\
1.85	0.7169\\
2.05	0.7101\\
2.25	0.7146\\
2.45	0.6817\\
2.65	0.7107\\
2.85	0.7199\\
3.05	0.6747\\
};
\addplot [color=blue, forget plot, line width = 1pt]
  table[row sep=crcr]{%
0.05	0\\
0.25	0.07009\\
0.45	0.08485\\
0.65	0.1738\\
0.85	0.321\\
1.05	0.2812\\
1.25	0.3222\\
1.45	0.4806\\
1.65	0.4474\\
1.85	0.5203\\
2.05	0.5038\\
2.25	0.5706\\
2.45	0.6133\\
2.65	0.5666\\
2.85	0.5727\\
3.05	0.5578\\
};
\addplot [color=green, forget plot, line width = 1pt]
  table[row sep=crcr]{%
0.05	0\\
0.25	0.1531\\
0.45	0.2628\\
0.65	0.2388\\
0.85	0.3749\\
1.05	0.4698\\
1.25	0.4307\\
1.45	0.5815\\
1.65	0.5281\\
1.85	0.5992\\
2.05	0.5774\\
2.25	0.6674\\
2.45	0.6299\\
2.65	0.6348\\
2.85	0.7212\\
3.05	0.6657\\
};
\addplot [color=red, line width = 1pt]
  table[row sep=crcr]{%
0.05	0\\
0.25	0.06268\\
0.45	0.1618\\
0.65	0.2628\\
0.85	0.3435\\
1.05	0.3143\\
1.25	0.5246\\
1.45	0.5846\\
1.65	0.5741\\
1.85	0.7169\\
2.05	0.7101\\
2.25	0.7146\\
2.45	0.6817\\
2.65	0.7107\\
2.85	0.7199\\
3.05	0.6747\\
};
\addlegendentry{$T_{60} = 0.2\,\mathrm{s}$}

\addplot [color=blue, line width = 1pt]
  table[row sep=crcr]{%
0.05	0\\
0.25	0.07009\\
0.45	0.08485\\
0.65	0.1738\\
0.85	0.321\\
1.05	0.2812\\
1.25	0.3222\\
1.45	0.4806\\
1.65	0.4474\\
1.85	0.5203\\
2.05	0.5038\\
2.25	0.5706\\
2.45	0.6133\\
2.65	0.5666\\
2.85	0.5727\\
3.05	0.5578\\
};
\addlegendentry{$T_{60} = 0.4\,\mathrm{s}$}

\addplot [color=green, line width = 1pt]
  table[row sep=crcr]{%
0.05	0\\
0.25	0.1531\\
0.45	0.2628\\
0.65	0.2388\\
0.85	0.3749\\
1.05	0.4698\\
1.25	0.4307\\
1.45	0.5815\\
1.65	0.5281\\
1.85	0.5992\\
2.05	0.5774\\
2.25	0.6674\\
2.45	0.6299\\
2.65	0.6348\\
2.85	0.7212\\
3.05	0.6657\\
};
\addlegendentry{$T_{60} = 0.6\,\mathrm{s}$}

\end{axis}
\end{tikzpicture}%
\vspace*{-3mm}       

    \caption{Output from the MRF-based detector for incremental shifts of a random node and varying $T_{60}$ with 100 trials per shift and $T_{60}$.}
    \label{MRF_localRes}
\end{figure}

For the results in Fig.~\ref{MRF_localRes}, the movement detection scenario was simulated over 100 trials per shift of a randomly chosen node, shifted in a random direction, and with random rotation, and for a range of reverberation levels. Obviously, the movement detection probability increases with the size of the displacement, and is largely independent of the $T_{60}$ level. We attribute the fact that the curves are not monotonic to the random rotation of the randomly moved node (from 0\degree - 360\degree).   

In order to test the robustness of the proposed algorithm, we compare it to a na\"{i}ve detector that uses the LONO positional estimates directly as a way of indicating movement. Thus the na\"{i}ve detector will indicate movement occurred if the deviation for a given LONO sub-network is greater than the average of the other three LONO estimates and this difference exceeds some threshold. Thereby the thresholds were chosen based on the average difference between a LONO sub-network excluding the shifted node and the mean estimates of the other three. We found (as indicated in Table~\ref{tab:AUC_comp}) that the MRF-based detector performs better for most $T_{60}$ levels with respect to the area under the curve (AUC) \cite{APB_ROCandAUC}. Note that the basis of the decision of the na\"{i}ve detector (i.e., $\mathbf{e}$), is essentially the input to the MRF-based detector. Thus, for the most part we observe improvement achieved by incorporating the prior information regarding the error distributions rather than the mean of the errors.
\begin{table}[htbp]
    \centering
\vspace*{-3mm}

    % \resizebox{\columnwidth}{!}{
    \begin{tabular}{p{2cm}p{1cm}p{1cm}p{1cm}}
    \\
        \toprule
               ${T_{60}}$~[s]&\multicolumn{1}{c}{{0.2}}&\multicolumn{1}{c}{{0.4}}&\multicolumn{1}{c}{{0.6}}\\ \midrule
        {Na\"{i}ve}&\multicolumn{1}{c}{0.71}&\multicolumn{1}{c}{0.62}&\multicolumn{1}{c}{0.78}\\
        {MRF}&\multicolumn{1}{c}{0.84}&\multicolumn{1}{c}{0.82}&\multicolumn{1}{c}{0.78}\\
        
        \bottomrule
    \end{tabular}
    % }
\vspace*{-1mm}       
    
    \caption[Source Localization Based On Semi-Supervised Approach]{AUCs reported for the LONO sub-network estimation comparison and the MRF-based detector at varying $T_{60}$.}
    \label{tab:AUC_comp}
\end{table}
 \vspace*{-3mm}

\section{Conclusion} \label{conclusion}

In this paper, we proposed a method for consistently identifying situations where moving sensor network nodes render source localization estimates questionable or useless. Specifically we considered the problem of detecting the movement of a microphone node in a network. The proposed probabilistic MRF-based algorithm determines whether a network of nodes is aligned with the previously learned configuration by leveraging prior information on the error distribution of an SSGP localization technique. The benefit of the MRF model was demonstrated by comparison to an estimate that relied directly on the relative difference in positional estimates by different sub-networks of nodes. In particular, we showed that the MRF-based detector outputs a movement indicator that scales commensurate with the size of disruption in the network, and one that is consistent across varying $T_{60}$. As of now the algorithm assumes a static source, and its application to a moving sound source is planned as future work.

\bibliographystyle{ieeetr}
\bibliography{bibli}
\end{document}